**Title**

**Thermoelectrically Elevated Hydrogel Evaporation for Personal Cooling under Extreme Heat**

**Authors**

Yu Pei [1]†, Tianshi Feng[1]†, Robert Chambers[1], Shengqiang Cai[1,2,] *, Renkun Chen[1,2,] *

**Affiliations**

[1]Department of Mechanical and Aerospace Engineering, University of California San Diego; 9500 Gilman Drive, La Jolla, CA 92093, USA.

[2]Materials Science and Engineering Program, University of California San Diego; 9500 Gilman Drive, La Jolla, CA 92093, USA.

*Corresponding author. Email: s3cai@ucsd.edu (S.C.); rkchen@ucsd.edu (R.C.)

†These authors contributed equally to this work

**Abstract**

Extreme heat events with wet-bulb temperatures (WBT) above 35°C pose serious risks to human survival, and conventional hydrogel evaporative cooling alone may not provide sufficient relief as it must be maintained at a sufficiently high temperature to achieve effective evaporation in hot, humid conditions. This study integrates thermoelectric devices (TEDs) with hydrogels to create an effective personal cooling solution. TEDs pump heat away from the skin to maintain comfort while simultaneously increasing the hydrogel's temperature to enhance evaporation. This hybrid system outperforms TEDs or hydrogel alone in extreme conditions (temperature up to 55°C and relative humidity up to 88%, with WBT > 35°C) and can operate for over six hours with a manageable hydrogel and battery weight. The active temperature control of TEDs allows adaptation to changing thermal loads and environments. These results demonstrate the potential of hybrid evaporative and thermoelectric cooling as an efficient, adaptable, and sustainable personal cooling solution to combat extreme heat.

**Teaser**

A hybrid cooling system combining thermoelectric and hydrogel enhances personal comfort in extreme heat conditions of WBT greater than 35 °C.

## MAIN TEXT

### Introduction

In recent years, extreme heat events have become more frequent, intense, and prolonged due to climate change and the urban heat island effect (*1*).The adverse health effects of human heat stress and the health risks associated with prolonged exposure to high temperature have become pressing concerns as global warming continues (*2, 3*). A key metric for assessing heat stress is wet-bulb temperature (WBT), which accounts for both temperature and humidity by representing the lowest temperature an object can reach through evaporative cooling (*4*). Unlike dry-bulb temperature (DBT), which is the temperature commonly reported in weather data, WBT provides a direct measure of the body's ability to dissipate heat through sweating. When WBT exceeds 28°C (82.4 F) , the risk of heat exhaustion increases significantly (*5, 6*). At WBTs above 35°C (95°F) for extended periods, human survival becomes extremely difficult, even for acclimatized individuals resting in the shade with unlimited access to water (*7*). Due to climate change, an increasing number of regions worldwide are experiencing WBTs exceeding 35 ºC, posing severe threats to human health and survivability. **Fig. 1** presents iso-WBT curves alongside data points representing extreme heat stress conditions, specifically the highest recorded global WBT values around the world (~36°C, shown in red circles). Such extreme WBTs have been reported in Centro, Mexico (DBT ~40°C, relative humidity (RH) ~75%), Hisar, India (DBT ~43°C, RH ~65%), and Dammam, Saudi Arabia (DBT ~50°C, RH ~40%). Meanwhile, heat-related illnesses account for more than 650 deaths and 65,000 hospitalizations annually in the United States alone, causing more fatalities than all other natural disasters combined (*8-10*). These issues are especially severe in regions lacking sufficient nighttime cooling or adequate infrastructure (*11*). As a result, there is an urgent need for advanced cooling technologies to mitigate the health risks posed by extreme heat.

Among various cooling strategies, personal wearable cooling devices and garments are gaining attention for their ability to directly cool the skin, providing thermal comfort on the go. This is particularly beneficial for those working outdoors in extreme heat or for individuals without access to air conditioning (*12*). Radiative personal cooling device increased the infrared (IR) radiation of the human body to passively dissipate heat utilizing materials with high emissivity in the atmospheric transparency window (8-13 µm) (*13-16*). While radiative wearable cooling can effectively dissipate heat in certain conditions, its performance is limited in environments with limited view factor to the sky or when the environment is too hot. Evaporative cooling, where liquid transforms into vapor by absorbing heat due to its high latent heat, is a powerful method of thermal regulation that we encounter frequently in daily life (*17*). The evaporation of sweat plays a key role in the body's natural heat dissipation of the body in hot environments (*18-20*). Consequently, there has been extensive research into easy-to-use evaporative cooling systems for personal thermal regulation. Moisture-responsive textiles that enhance vapor permeability when activated have shown notable cooling performance compared to standard fabrics(*21, 22*). However, these textiles depend on sweat evaporation from the skin and are therefore less effective in extreme heat exceeding 40°C, as sweating at such high temperatures can lead to discomfort.

Alternatively, evaporative cooling can be achieved through textiles that contain water. In particular, hydrogel—a material composed of gel networks with a high water content—offer a promising solution due to their ability to store water and release it through evaporation. Hydrogels can be easily applied as coatings on textiles, and maintains its elasticity across a wide range of temperatures.(*23, 24*) Recently, Kim et al. reported the development of highly entangled, strong, and wear-resistant hydrogels with potential applications in effective evaporative cooling.(*25*) Hydrogel enhanced thermal management capability has been demonstrated in various electronic systems (*26, 27*) and holds potential for personal thermal management applications. For instance, Fei et al. developed a hydrogel-based composite capable of providing sub-ambient cooling power of 350 W m$^{-2}$ in environments up to 36°C (*28*). Later, Hu et al. demonstrated a continuous cooling effect lasting up to an hour under metabolic heating in ambient temperatures averaging 43°C, using a hierarchically porous hydrogel composite (*29*)**.** Li et al. report a method for spray-coating porous hydrogel coatings over large areas which can enhance hydrogel cooling efficiency (*30*). However, the operational temperature limit remains below 45°C, insufficient to counter extreme temperatures above 50°C observed in many regions around the globe (*31-33*).

One limitation is the low surface temperature of the hydrogel, which affects the evaporation rate. When the skin temperature is maintained at a comfortable level around 35°C, the evaporation surface temperature of the hydrogel is determined by its thermal resistance and metabolic heat generation rate. Typically, with a 5 mm thick hydrogel, there is a temperature drop of about 2°C across the hydrogel layer (**Fig. 2D**), resulting in a surface temperature of around 33°C. This temperature is insufficient to drive rapid evaporation in the hot and humid conditions of extreme heat, thereby reducing the effectiveness of hydrogel in such environments.

To address the limitations of current hydrogel-based personal cooling devices in extreme heat, we propose a hybrid wearable cooling garment that combines hydrogel with thermoelectric devices (TEDs), as shown in **Fig. 2, A to C**. Thermoelectric (TE) cooling utilizing the Peltier effect has been widely explored as personal cooling solutions due to their lightweight, solid-state structure, and active controllability (*34-37*). TEDs use electric current for heat pumping across a junction of two dissimilar materials, enabling localized cooling, making them ideal for applications in wearable technologies. However, effectively dissipating heat from the hot side of TEDs to a hot environment is a significant challenge that directly impacts their performance and efficiency. Insufficient heat dissipation reduces the temperature differentials across the device, thereby diminishing its cooling capacity. Common methods for TED heat dissipation include liquid coolant loops (*38, 39*) and circulating air (*40*), but these approaches are either bulky and heavy or incapable of providing sufficient cooling at higher ambient temperatures. Additionally, previous research has primarily focused on developing wearable TEDs for small, limited functional areas (typically less than 5x5 cm²) (*41-44*). Therefore, a large-area, lightweight TED system with effective heat dissipation mechanisms is needed to fully realize the potential of TED-based wearable cooling.

Building on our previous work (*45*) in optimizing thermal design to establish an ideal temperature gradient across TEDs, we combine TEDs and hydrogel to effectively elevate the surface evaporation temperature of the hydrogel on the TEDs' hot side, while maintaining a comfortable temperature on the cold side. This TED-hydrogel composite is integrated into a fabric, creating a wearable active cooling garment designed to perform under extreme heat conditions (**Fig. 2**). During operation, the hydrogel absorbs heat from the hot side of the TED, initiating sufficient evaporation to dissipate both metabolic heat and excess TED-generated heat. As shown in **Fig. 2D**, the TED-hydrogel device achieves a higher surface temperature in the hydrogel, resulting in a more efficient evaporative cooling rate than hydrogel alone. Consequently, this TED-hydrogel integration enables more effective cooling of the skin in the same ambient conditions.

We evaluated the impact of hydrogel surface temperatures on evaporation rates through evaporation rate testing. Our findings show the importance of high surface temperatures for achieving optimal evaporation. COMSOL simulations further validated that the TED-hydrogel composite significantly exceeds the cooling rates achievable by hydrogel alone. Testing in controlled environments, with ambient temperatures from 40°C to 55°C and relative humidity levels of 30%-88%, demonstrated that the device consistently maintained skin temperatures within the comfort threshold of 35.8°C (*17*) across all tested conditions, even at an ambient temperature of 55°C. The adaptability of the system was further highlighted by its response to varying thermal loads typical of different activities, ranging from 50 to 150 W m$^{-2}$. Long-term durability tests showed that a garment with 5 mm thick hydrogels effectively operated for over six hours. Integrating a temperature controller with the TED-hydrogel device ensured stable skin temperatures when subjected to fluctuatingenvironmental conditions, showcasing the benefits of dynamic temperature regulation over traditional passive cooling methods.

Our study demonstrates that the TED-hydrogel integration offers a novel approach to personal cooling, meeting the demand for more effective and adaptable personal cooling solutions amid global temperature rises. With a sustained performance over several hours, this system is well-suited for outdoor workers and those vulnerable to heat-related illnesses in increasingly hot climates.

**Results**

**Hydrogel Evaporation rate**

The evaporation rate of the hydrogel is closely linked to the ambient environment and its surface temperature. To investigate this, we conducted a mass loss test. **Fig. 3A** shows the experimental setup, where a 4 cm × 4 cm × 5 mm hydrogel sample was placed on a balance in an environmental chamber with controlled temperature and humidity. After thermal equilibrium, the hydrogel surface temperature was measured using an IR camera (**Fig. 3B**). Due to evaporative cooling, the hydrogel surface temperature was lower than the ambient temperature ($T_{amb}$), with the difference depending on both environmental temperature and humidity. For example, **Fig. 3B** shows that the hydrogel surface temperature is about 33 ºC when the ambient temperature is 40 ºC and the relative humidity is ~45% ($T_{amb}$= 40ºC, RH ~45%). At the same ambient temperature, lower RH led to a lower hydrogel surface temperature due to a higher evaporation rate. We calculated the mass loss ratio, $\frac{m_0-m(t)}{m_0}$, by recording the initial weight $m_0$ and real-time weight $m(t)$ over time. Mass loss was measured at two ambient temperatures (40°C and 50°C) and two humidity levels (RH = 30–35% and RH = 45–50%).

Evaporation from a surface is affected by ambient temperature ($T_{amb}$), air humidity, surface water temperature ($T_s$), and air velocity above the water surface ($v$). Here, we assume that the evaporation of hydrogels is the same as the evaporation of free water. The evaporation rate $g_s$ [kg s$^{-1}$ m$^{-2}$] can be described by the equation(*26*):

$$g_s = (25 + 19v) * (X_s - X_{sair}) * 3600 \qquad [1]$$

where $X_s$ is the maximum humidity ratio of saturated air at the surface temperature, $X_{sair}$ is the humidity ratio of air at ambient temperature.

Further evaporation calculations are detailed in Supporting Materials. Calculation results for the mass loss ratio, represented by the shaded areas in **Fig. 3C**, show good agreement with the experimental data. The upper and lower limits of the shaded area were obtained by considering the hydrogel surface temperature range in the evaporation rate equation. Evaporation is driven by the vapor pressure difference between the water surface and the ambient. At a given ambient temperature, lower humidity means lower air water vapor pressure ($P_{air}$), resulting in a higher evaporation rate and lower hydrogel surface temperatures. While at a fixed RH, increasing the ambient temperature raises the hydrogel surface temperature, leading to a higher saturation vapor pressure of water in the hydrogel ($P_{sat,hydrogel}$), thereby increasing the evaporation rate.

**Fig. 3D** shows the calculated evaporation rate as a function of ambient and hydrogel surface temperatures, with the right z-axis converting the evaporation rate into the corresponding heat flux using the latent heat of evaporation of water. Experimental results in **Fig. 3C** show that the difference between ambient and hydrogel surface temperatures ($T_{amb} - T_s$) increases as $T_{amb}$ rises. Calculation results in **Fig 3D** indicate that when this difference is large, the evaporative cooling power of hydrogel is minimal, meaning that the cooling is diminished at higher ambient temperature. Additionally, when $T_{amb}$ is significantly higher than $T_s$, convection and radiation heat flux from the environment to the hydrogel becomes important, which may result in heating, rather than cooling, of the skin.

To evaluate the suitable conditions for hydrogel evaporative cooling, we used COMSOL simulations to estimate the cooling power of hydrogel under various temperature and humidity conditions (details in the experimental section). The bottom temperature of the hydrogel (representing target skin temperature) was set to 34°C. The calculated net cooling power provided by the hydrogel under different conditions is shown in **Fig. 3E**. For temperatures below 40°C, the hydrogel dissipates heat fluxes of 50–150 W m$^{-2}$ effectively, which represents the typical human metabolic heat flux. However, at ambient temperatures above 40°C and RH levels above 30%, the evaporation rate is limited by the hydrogel temperature. In these conditions, hydrogel evaporation cannot provide sufficient cooling to offset the metabolic heat. In fact, at ambient temperatures above 42°C and RH levels above 30%, the net cooling power becomes negative, indicating that radiation and convection heat flux from the hotter environment overwhelms the evaporation cooling flux.

To overcome the limitation of inadequate evaporation when the hydrogel temperature is below ambient temperature, especially at higher humidity levels, we propose a TED-hydrogel tandem device consisting of a hydrogel layer placed on top of the hot side of the TED while the cold side of the TED is in contact with the skin . As shown in **Fig. 2D**, the TEDs not only add cooling capacity to the skin but also elevates the hydrogel temperature to achieve a high evaporation rate. The TED design was selected based on our previous work (*46*), which was optimized to establish a large temperature difference between the hot and cold sides and is thus suitable for the purpose of elevating the hydrogel temperature.

We developed a COMSOL model to evaluate the system performance, considering hydrogel evaporation, thermoelectric effects, and heat transfer between the device and the environment. In the simulation, we fixed the skin surface temperature at 34°C because it closely reflects real-world applications. Human skin temperature is generally maintained between 33°C and 36°C through thermoregulation (*47*). The TED-hydrogel personal cooling garment is designed to provide adjustable cooling power, ensuring that the skin remains within a comfortable temperature range. As shown in **Fig. 3E**, simulations of the TED-hydrogel hybrid demonstrate significant improvements in cooling capacity, even at ambient temperatures up to 55°C and 50% RH. The net cooling power for the TED-hydrogel device shown in **Fig. 3E** is achieved under a specific applied voltage of 0.6 V on the TED. One can adjust the applied voltage to control the net cooling power to match with the human metabolic heat under specific ambient conditions. For example, fig S1 (Supplementary Materials) shows COMSOL simulation results of net cooling power in the range of 60-300 W m$^{-2}$ for the applied voltage from 0 to 0.3 V, under DBT = 40°C and RH = 40% (WBT = 28.5°C). In the simulated temperature distribution (**Fig. 3F**), while the bottom surface of the TED-hydrogel tandem remains at 34°C, the top surface of the hydrogel reaches up to 50°C due to the heat pumping of the TEDs, significantly enhancing the evaporation rate of water in hydrogel. fig. S2 (Supplementary Materials) shows more simulation results of temperature distribution in the hydrogel and the TED-hydrogel devices under different environmental conditions.

**Single TED-hydrogel cooling device**

To experimentally test our concept, we first used a single TED (2.3 × 2.3 cm) in combination with a 2.5 cm × 2.5 cm × 5 mm hydrogel sample. A heater was placed beneath the TED to simulate the metabolic heat flux from human skin. When current was applied to the TED, the bottom surface of the TED cooled while the top surface heated, with the combined heat from the TED and the heater being dissipated through hydrogel evaporation. This setup was placed in an environmental chamber with controlled temperature and humidity. **Fig. 4A** shows a schematic of the experimental setup, where the heater was connected to a power supply, and the temperatures on the hot and cold sides of the TED were measured using thermocouples connected to a data acquisition (DAQ) module (HP HEWLETT PACKARD 34970A) for data recording. The TED was connected to a voltage source (Agilent E3634A) controlled by a computer to adjust the applied voltage. **Fig. 4B** shows an image of the experimental TED-hydrogel device.

Furthermore, we have conducted modeling and experiments for different DBT and RH combinations corresponding to the same WBT of 38°C, including DBT ~40 °C & RH ~88%, DBT ~42 °C & RH ~77%, DBT ~45 °C & RH ~63%, and DBT ~50 °C & RH ~45% (as shown in the diamond mark in **Fig.1**), which simulate even more severe conditions than those previously recorded in Centro, Mexico and Hisar, India. Simulation results in fig. S3 (Supplementary Materials) demonstrates that these different DBT-RH combinations result in the similar evaporative cooling rate (~300 W m$^{-2}$) from our hydrogel + TED system due to the same WBT, which is higher than the typical human metabolic rate range for activities such as resting, standing, light activities, and walking at 3 km/hr. Since evaporative cooling—whether from human sweating or hydrogel—depends on WBT, different combinations of RH and DBT can result in the same cooling effect as long as the WBT is the same. **Fig. 4C** shows the results for the cold and hot sides of the TED-hydrogel device under the same wet-bulb temperature (~38 °C). Despite the different DBT values, the TED+Hydrogel systems showed similar cooling performance and skin temperature ($T_c$) due to the similar WBT values. For comparison, we also conducted an experiment with hydrogel placed directly on the heater (i.e., hydrogel only). As shown in **Fig. 4D**, while the TED+hydrogel devices have the ability to provide sufficient cooling to the skin (i.e., below 35 °C), the hydrogel-only system fails to provide adequate cooling at such high WBT, as the hydrogel surface temperature is not high enough to sustain high evaporation rates in such hot conditions.

**Fig. 4E** present the experimental results under different ambient temperatures (40°C, 45°C, 50°C, and 55°C) and 30–35% RH, with heat flux of 100 W m$^{-2}$ on the heater. The voltage applied to the TED ranged from 0.1 to 1V. As the voltage increased, the cold side temperature ($T_c$) of the TED—representing skin temperature—decreased, while the hot side temperature ($T_h$) increased. Both $T_c$ and $T_h$ rose as the ambient temperature increased. Similar experimental results under 45–50% RH is shown in fig. S4 (Supplementary Materials). The results for comparing different ambient humidty condition showing in **Fig. 4F**, we can also conclude that higher humidity levels led to higher $T_c$ and $T_h$ at the same ambient temperature, as the hydrogel evaporation rate decreases at higher humidity.

Long-term stability tests on the TED-hydrogel and hydrogel-only cooling device are conducted at a constant applied voltage of 0.6V under varying temperature and humidity conditions, the results displayed in fig. S5, A to C (Supplementary Materials). The hydrogel-only case has about 5 °C higher cold side temperature compared to that of TED-hydrogel under all testing conditions. In particular, at ambient temperature of 45 °C and RH of 45-50 % (fig. S5B) or ambient temperature of 50 °C and RH of 30-35 % (fig. S5C), the hydrogel only case had cold side temperature of about 37.5 °C, above the the skin thermal comfort range. In all scenarios, the TED-hydrogel tandem device maintained the skin surface temperature below 31°C, with stable performance for over an hour.This result again demonstrates that the TED-hydrogel device can provide adequate cooling for human skin under extreme conditions where hydrogel alone cannot.

**TED-hydrogel Personal Cooling Garment**

After confirming the feasibility of the TED-hydrogel device for personal cooling, we designed a personal cooling garment, shown in **Fig. 5A**. The garment consists of a 4 × 4 array of TEDs attached to a fabric with high thermal conductivity (Dyneema, $k$ ~ 14.2 to 28.4 W m$^{-1}$ K$^{-1}$)(*46*), with each TED covered by a hydrogel layer on the top to dissipate heat. The spacing between TEDs allows the garment to remain flexible and comfortably conform to the curvature of the human back. **Fig. 5A** illustrates the cross-section of the cooling setup.

In addition to measuring the hot side temperature ($T_h$) and cold side temperature ($T_c$) of the TEDs, we monitored the fabric temperature at the middle of the TED array ($T_m$), representing the highest temperature on the fabric due to the lateral thermal resistance of the fabric. The garment was tested in an environmental chamber under different temperatures (45°C and 50°C) and humidity conditions (30–35% RH and 45–50% RH). The TED array was wired into two parallel strings, each with eight TEDs in series. The total applied voltage ranged from 1–9V. As shown in **Fig. 5B**, the garment effectively maintained $T_c$ below 35°C in low-humidity conditions (30–35% RH) at both 45°C and 50°C ambient temperatures. In higher humidity (45–50% RH), cooling performance decreased due to a lower evaporation rate, though $T_c$ remained within the thermal comfort range at both temperatures, as shown in fig. S6.

To simulate various human activities with different metabolic heat rates (e.g., resting or sitting at ~45–55 W m$^{-2}$, slow walking at ~110 W m$^{-2}$, and fast walking at ~140 W m$^{-2}$) (*47*),we conducted additional tests with varying heater power levels. The data for skin temperature $T_c$ and hot side temperature $T_h$ of the TED-hydrogel device are shown in fig. S7 (Supplementary Materials). As shown in **Fig. 5C**, the TED-hydrogel cooling garment performed effectively at 50°C ambient temperature and 30–35% RH across a range of simulated body heat flux levels up to 150 W m$^{-2}$.

We also conducted long-term tests with hydrogels of different thicknesses (3 mm and 5 mm). The 5 mm thick hydrogel maintained $T_c$ below 35.8°C for over 6 hours, while the 3 mm thick hydrogel lasted for 4 hours. As shown in **Fig. 5D**, with a fixed applied voltage of 5 V to the TEDs, $T_c$ initially as low as ~28°C, and $T_h$ is ~47.5°C due to the heat pumping of the TED to its hot side, and the temperatures remained stable for approximately 100 minutes. Subsequently, all temperatures gradually rose as the hydrogel losing water, causing it to shrink laterally and reducing cooling power due to decreased evaporating surface area. In the real-world application, $T_c$ can be maintained at a constant temperature by adjusting the applied voltage to the TEDs with a temperature controller, as shown in our subsequent experiment (**Fig. 6**). Eventually, the hydrogel surface area became smaller than the TED, causing the edges of the hydrogel to peel off and resulting in a rapid rise in both $T_c$ and $T_h$. Throughout the test, the temperature at the middle of the fabric ($T_m$) remained a few degrees higher than $T_c$ due to the lateral thermal resistance of the fabric.

The initial weight of the 16 pieces of 5 mm hydrogel was 154.46 g, which decreased to 58.55 g after 7 hours, reflecting a 61% mass loss. Fig. S8 (Supplementary Materials) shows the changes in 5mm thick hydrogel size over test time. The average evaporation rate, calculated as $\frac{m_0 - m}{t * A}$, where $A$ is the initial area of the hydrogel and $t$ is time, was $1.6853 \times 10^{-4}$ kg s$^{-1}$ m$^{-2}$, matching the calculated value of $1.835 \times 10^{-4}$ kg s$^{-1}$ m$^{-2}$ (**Fig. 3D**). For the 3 mm hydrogel, the mass loss was 66% (from 93.78 g to 31.56 g) after 4.5 hours, with an average evaporation rate of $1.753 \times 10^{-4}$ kg s$^{-1}$ m$^{-2}$. We note that the hydrogel can be restored to its original size by soaking in water for 1–2 hours.

By assembling the TED-hydrogel cooling device with a PID (Proportional-Integral-Derivative) temperature controller and a battery pack, we tested the personal cooling garment in a real-world environment with varying temperature and humidity (**Fig. 6A**). The DC output from the battery pack was connected to the temperature controller as the power input, and the controller output supplied the power to the TEC array. The controller generates a fixed-period square wave with preset maximum and minimum voltage values, adjusting the duty cycle based on the temperature setpoints.

**Fig. 6B** shows IR images of the TED-hydrogel front (hot) and back (cold) sides of the cooling device when operating at an ambient temperature of 45°C and 45–50% RH. The back side maintained a temperature close to the setpoint (34°C), within the comfortable range for the human back. The top surface of the hydrogel reached a significantly higher temperature than the cold side of the TEDs, enabling an elevated evaporation rate for effective heat dissipation.

**Fig. 6 C and D** show the resulting $T_c$, $T_h$, and $T_m$ values with ambient temperature of 45°C under different humidities, with similar results at 40°C presented in fig. S9 (Supplementary Materials). The red line indicates the average output current from the battery pack. As shown, $T_c$ closely followed the temperature setpoint, while $T_m$ remained a few degrees higher but followed the same trend. As the temperature setpoint increased, both the current and $T_h$ decreased, due to the lower heat pumping and evaporation cooling need at higher $T_c$. At 45°C ambient temperature and 45–50% RH, $T_c$ began to deviate from the setpoint when the setpoint was around 30°C as the battery pack reached its maximum output capacity.

Recognizing that ambient temperatures in real-world applications may fluctuate, we conducted tests with ambient temperatures changing from 40–45°C at 45–50% RH. **Fig. 6E** shows the results of this dynamic test, with the temperature setpoint fixed at 34°C and the heater providing a heat flux of 100 W m$^{-2}$. Due to the tempeature control, $T_c$ remained stable at the set temperature while the ambient temperature changed from 40°C to 45°C. This is achieved by raising the $T_h$ to attain a higher evaporation flux from the hydrogel via increasing the power applied to the TED, as shown in the gray shaded area..

**Discussion**

We summarize the cooling performance of the TED only, hydrogel alone, and the TED-hydrogel combination in **Fig. 7**. The data for the TED only case is from our recent work (*46*), where the TED was cooled by forced convection of air using a fan. All three methods demonstrate notable cooling power compared to the baseline without any cooling device. The RH for all tests was maintained at approximately 35%. At an ambient temperature (DBT) of 40°C (corresponding to WBT of 28°C), compared to the no device case (black square), the skin temperatures with hydrogel, TED, and the TED-hydrogel were 15.9 °C, 12.33°C and 26.9°C lower respectively. The TED with air forced convection provides effective cooling at ambient temperatures below 40°C (WBT ~ 28 °C). However, above this temperature, convective heat dissipation alone cannot adequately remove the heat generated by the TED. In contrast, hydrogel alone shows better cooling performance than the TED alone at higher temperatures, though its cooling capacity becomes insufficient in environments above 45°C (WBT of 32°C). Among the three methods, the TED-hydrogel device shows the best cooling performance under extreme heat conditions, with WBT up to 40 °C (DBT up to 55 °C), as the TED contributes to the cooling of the skin on its cold side while its hot side elevates the hydrogel temperature for higher evaporation rate. This synergy between the TED and hydrogel accounts for the superior performance of the TED-hydrogel combination shown in **Fig. 7**.

Hydrogel is a promising material for passive personal cooling due to its ability to dissipate heat through evaporation. However, its cooling power alone is limited by the relatively low evaporation rate under typical ambient conditions. Achieving a higher evaporation rate would require a consistently high hydrogel surface temperature, which would be too high for the skin. TED is a widely used cooling technology, but its effectiveness depends heavily on efficient heat dissipation from the hot side, typically achieved with a solid heat sink or fan. Our research demonstrates that hydrogel is a more effective heat dissipation medium for TEDs than conventional methods. The TED-hydrogel combination leverages the benefits of both systems: the TED provides active cooling and raises the temperature of hydrogel, while the hydrogel enhances heat dissipation through evaporation at elevated temperature.

Additionally, the TED-hydrogel device offers better control over skin temperature by adjusting the TED power output in response to environmental changes. In contrast, hydrogel alone varies in effectiveness with environmental conditions, potentially leading to skin temperatures that are either too high or too low for thermal comfort. This makes the TED-hydrogel combination a more reliable and adaptable solution for personal cooling across diverse environmental conditions.

For real-world application, the weight of the garment is important. The total weight of the garment studied here is approximately 530 g: ~154 g for 16 pieces of 5 mm thick hydrogel: 256 g for the battery pack, 88 g for the 16 TEDs, and ~31 g for the temperature controller. There is a correlation between the system weight and the cooling duration, primarily due to the capacity of the battery pack and the hydrogel. The The battery pack shown here has a capacity of 59.2 Wh (16,000 mAh/3.7V). Based on the maximum power consumption of ~2.5 W at 45 °C ambient temperature, the battery can operate the garment for about 24 hours, more than sufficient to sustain the cooling duration of the 5 mm thick hydrogel (~ 6 hours). If a lighter garment is desirable, one could opt to use a smaller battery pack and thinner hydrogel. For example, with the 3 mm hydrogel (original weight of ~94 g) and 1/6 battery pack (~43 g weight), the cooling duraiton would be about 4 hr and the the garment weight would be about 256 g.

In conclusion, this study demonstrates that integrating thermoelectric devices (TEDs) with hydrogels provides a powerful and adaptable personal cooling solution that outperforms each component alone, particularly under extreme environmental conditions (up to 55°C in temperature and 88% in RH, or WBT > 35°C ). By actively raising the hydrogel surface temperature, TEDs enhance the evaporative cooling effect of hydrogel, delivering stable cooling in both constant and fluctuating temperatures and humidity levels representative of extreme heat. Experimental results show that the TED-hydrogel combination can maintain a skin-like surface temperature below the thermal comfort threshold of 35.8 °C for extended periods (over 6 hours with a 5 mm hydrogel layer), significantly increasing the evaporation rate and heat dissipation. In real-world applications, the flexible TED-hydrogel personal cooling garment not only improves cooling efficacy but also enhances adaptability, supporting sustainable thermal comfort in extreme heat. This research paves the way for the advancement of personal cooling technologies to combat extreme heat conditions that are increasingly intensified by climate change.

**Materials and Methods**

**COMSOL Model of Hydrogel Evaporation**

The thermal performance of hydrogel evaporation was simulated using the COMSOL Multiphysics Heat Transfer Module, with hydrogel thermal properties assumed to be the same as water. The bottom surface of the hydrogel was set at a constant temperature of 34 °C. Convection, radiation, and evaporation boundary conditions were applied to the top and side surfaces, with the hydrogel emissivity assumed to be 0.9 under natural convection. The evaporation rate was determined by Equation [1], which estimates pure hydrogel evaporation as a function of hydrogel surface temperature, ambient temperature, and relative humidity. A stationary study was conducted to simulate the experimental steady state, and the cooling power was extracted as the total normal heat flux at the bottom surface.

**COMSOL Model of the Hybrid System (TED+Hydrogel)**

The TED performance was simulated by combining the thermoelectric effect and electromagnetic heating modules in COMSOL. The model geometry was constructed based on the device size and its 36 thermoelectric (TE) pillars, as well as the air layer between two alumina substrates. A constant temperature boundary condition of 34 °C was applied beneath the bottom alumina substrate. The top of the alumina substrates was covered with a hydrogel layer, serving as an additional heat sink. Radiation and convection boundary conditions were applied to all surrounding and top surfaces, with the same evaporation rate boundary condition as in the hydrogel-only model. Using a stationary study with an input current of 250 mA, the cooling power was calculated by extracting the total heat flux at the bottom TED surface.

**Materials**

All chemicals were used as received without further purification. Acrylamide (AAm; Sigma-Aldrich 01700), N,N'-methylenebis(acrylamide) (MBAm; Sigma-Aldrich M7279), Ammonium persulfate (APS; Sigma-Aldrich A3678), and N,N,N',N'-tetramethylethylenediamine (TEMED; Sigma-Aldrich T9281) were used in hydrogel synthesis.

**Synthesis of PAAm Hydrogel**

Polyacrylamide-based hydrogel synthesis involved the polymerization of acrylamide (AAm) monomers, cross-linked by N,N′-methylenebis(acrylamide) (MBAm). Ammonium persulfate (APS) served as the initiator, generating free radicals to start polymerization, while tetramethylethylenediamine (TEMED) acted as a catalyst to accelerate polymerization by promoting radical formation from APS. After combining all components, the polymerization formed a flexible, hydrophilic three-dimensional hydrogel network. Hydrogel synthesis methods were adapted from previous work (*41*). Three solutions and TEMED were mixed to form the polymer precursor, which was then poured into the mold. Solutions were separately prepared and stirred with a magnetic stirrer until homogenous. Solution (1) was prepared with 25 wt% monomer, combining 30 g AAm and 90 g deionized water. Solution (2) consisted of 10 g deionized water and 0.2 g MBAm (2 wt%). Solution (3) contained 10 g deionized water and 0.5705 g APS (0.25 M). The mixing ratio was 1 g of solution (1), 9 µL of solution (2), 8 µL of solution (3), and 0.8 µL TEMED. For a batch using 120 g of solution (1), 1080 µL of solution (2), 960 µL of solution (3), and 96 µL of TEMED were added to solution (1) while stirring.
The mold consisted of a 4 × 4 cavity array. A laser-cut acrylic sheet was adhered to a glass plate using VHB to create a watertight mold. The upper glass plate, cavities, and lower glass plate were treated with RainX to facilitate release. After preparing the precursor solution, a transfer pipette was used to fill the mold, slightly overfilling each cavity. The upper glass plate was then lowered carefully and secured with weights. The mold was held at room temperature for 1 hour to partially cure the hydrogel, followed by 2 hours in a 45°C oven. Once cured, the hydrogel was demolded.

**Acknowledgments**

**Funding:**

**Author contributions:**
    Conceptualization: RC, YP, TF
    Methodology: RC, YP, TF
    Investigation: YP, RC
    Visualization: YP, TF
    Supervision: RC, SC
    Writing—original draft: RC, YP, TF, RC, SC
    Writing—review & editing: RC, YP, TF, RC, SC

**Competing interests:**
All other authors declare they have no competing interests.

**Data and materials availability:**
All data are available in the main text or the supplementary materials.

**Figures and Tables**

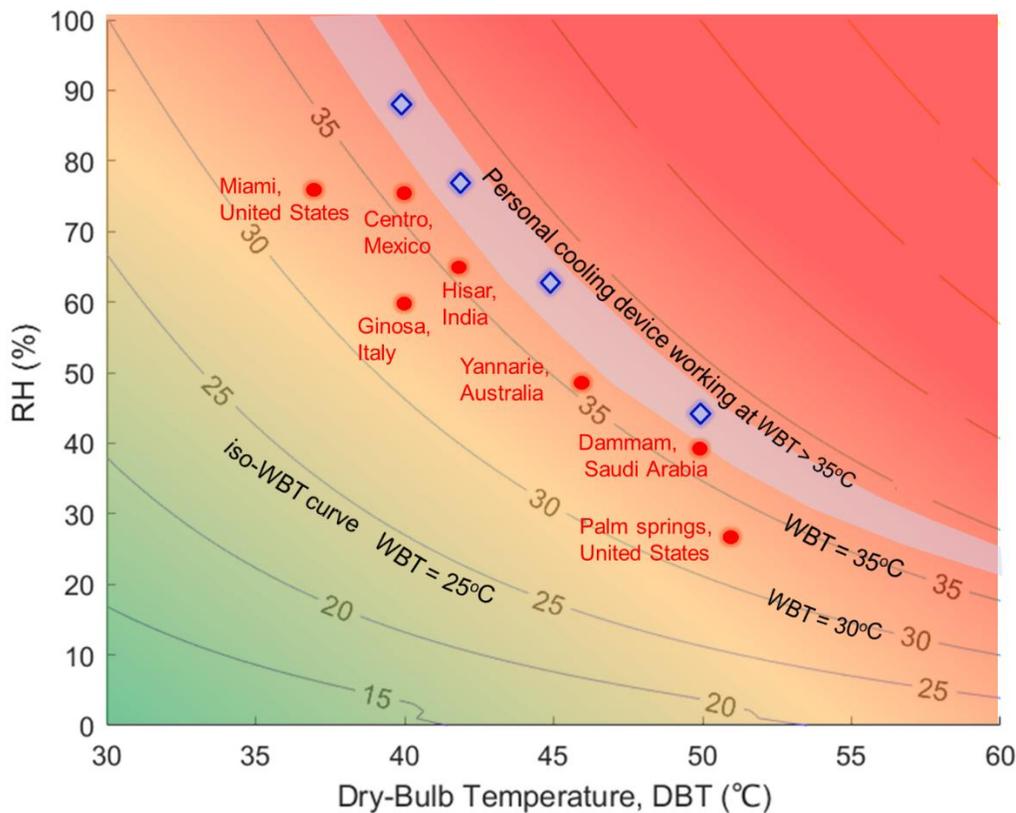

**Fig. 1. The iso-WBT (wet-bulb temperature) curves with different DBT (dry-bulb temperature) and RH (relative humidity).** The red circles show the highest reported WBT (~36 °C) in different regions around the world. The diamond marks are some experimental conditions we used to verify that the device can work at WBT~ 38 °C.

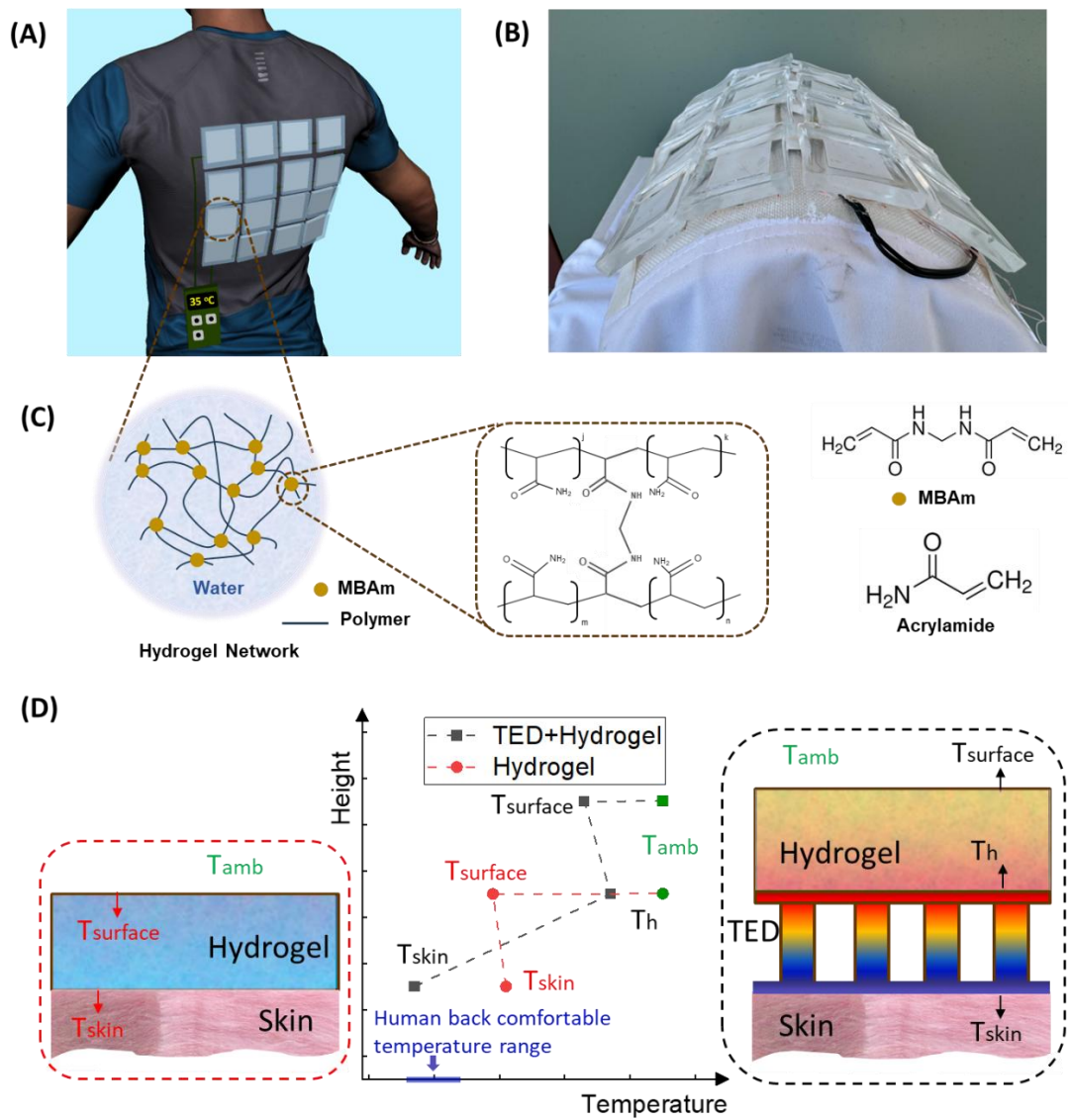

**Fig. 2. Novel personal cooling garment integrating thermoelectric devices and hydrogel (TED-hydrogel).** (**A**), Schematic of the TED-hydrogel personal cooling garment. (**B**), Photograph of the flexible, wearable TED-hydrogel active cooling garment. (**C**), Schematic and chemical formula of the polyacrylamide (PAAm) hydrogel, with methylenebisacrylamide (MBAm) as a cross-linker for acrylamide (AAm) monomers, forming a hydrogel network through polymerization. (**D**), Temperature distribution along the thickness direction for the hydrogel and TED-hydrogel device. $T_{skin}$ represents the interface temperature between the hydrogel and the skin, or between the cooled TED side and the skin. $T_h$ denotes the hot side temperature of the TED, $T_{surface}$ is the top surface temperature of hydrogel, and $T_{amb}$ is the ambient temperature. The shaded blue bar on the x-axis represents the comfortable temperature range for the human back.

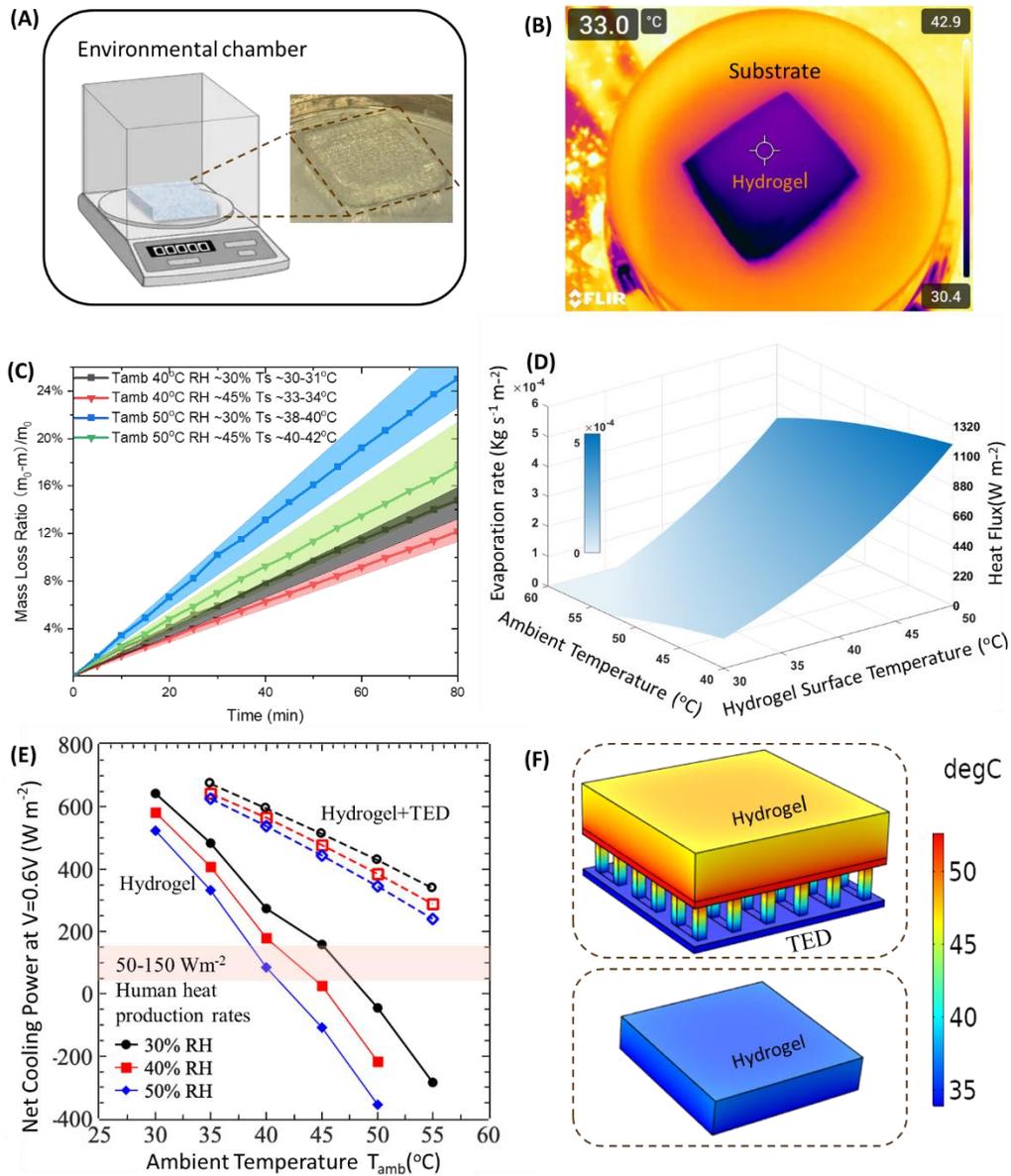

**Fig. 3. Hydrogel evaporation rate and cooling power.** (**A**), Setup for measuring hydrogel mass loss. (**B**), Infrared (IR) image showing the hydrogel surface temperature ($T_s$) during evaporation (ambient temperature $T_{amb}$ = 40 °C and relative humidity RH ~45%). The purple region represents the hydrogel, which has a lower temperature than the ambient due to evaporative heat dissipation. (**C**), Measured (lines and symbols) and calculated (shaded areas) hydrogel mass loss ratio. The mass loss ratio is defined as the mass loss $m_0 - m(t)$ divided by the original hydrogel weight $m_0$. Calculations used the range of $T_s$, as shown in the legend. (**D**), Calculated hydrogel evaporation rate under different ambient temperatures (with RH fixed at 30%) and hydrogel surface temperatures. The right z-axis represents the corresponding heat flux from water evaporation, which is the mass of evaporated water multiplied by latent heat. (**E**), COMSOL simulation results showing net cooling power as a function of ambient temperature for hydrogel alone (symbols + solid lines) and hydrogel + TED (symbols + dashed lines) under 0.6V applied voltage to the TED. The shaded band represents the metabolic rate. Net cooling power includes evaporative cooling minus the heat transfer rate from ambient to skin, with skin temperature fixed at 34°C. The net cooling power of the TED-hydrogel device can be adjusted with the applied voltage (fig. S1, Supporting Information). The negative net cooling power of the hydrogel-alone case means the heat gain from the hot environment outweighs the evaporative cooling from the hydrogel. (**F**), Temperature distribution for hydrogel alone and TED-hydrogel from the COMSOL model at $T_{amb} = 55 °C$, RH = 30%, and net cooling power of 338 W m$^{-2}$ under 0.6 V applied voltage for the TED-hydrogel device and -284 W m$^{-2}$ for hydrogel alone.

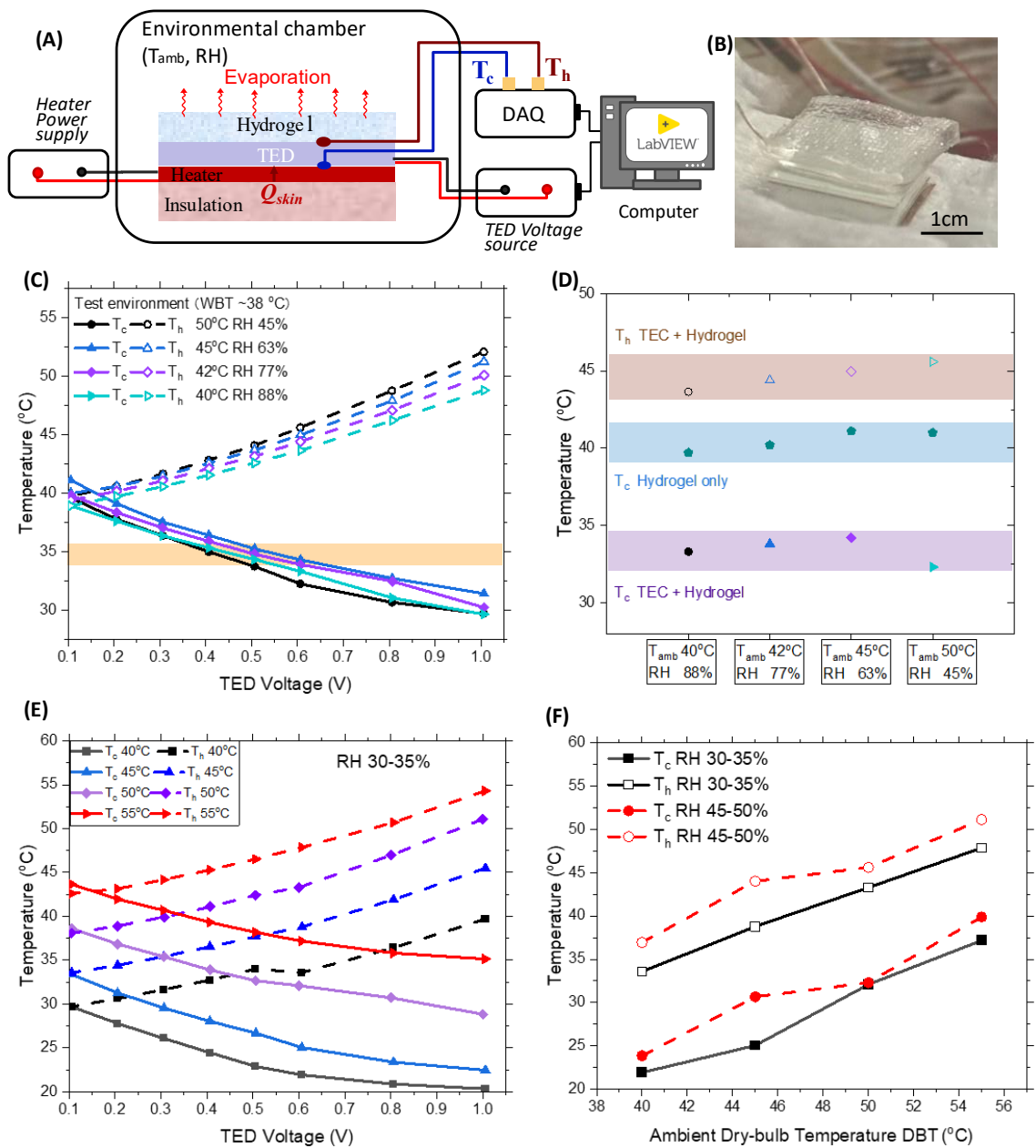

**Fig. 4. Performance of a single TED-hydrogel cooling device.** (**A**), Measurement setup for the TED-hydrogel device. The power supply delivers a heat flux of 100 W m$^{-2}$ to the heater to simulate human metabolic heat flux. The computer controls the TED voltage supply and records the temperatures of the cold ($T_c$) and hot ($T_h$)

sides of the TED. (**B**), Photograph of the single TED-hydrogel cooling device. (**C**), Experiment results for the cold and hot sides of the TED-hydrogel device under same wet-bulb temperature (~38 °C) but different ambient temperatures and humidity conditions. Solid lines represent the cold side ($T_c$) and dashed lines represent and hot side ($T_h$) of the TED at various TED voltages. The power supply delivers a heat flux of 100 W m$^{-2}$ to the heater to simulate human metabolic heat flux. (**D**) Performance of TED-hydrogel and hydrogel-only systems under various dry-bulb temperature (DBT) and relative humidity (RH) values, but with the same WBT of ~38 °C. (**E**) Experiment results for the cold and hot sides of the TED-hydrogel device under different ambient temperatures and 30–35% RH. (**F**) Comparation of $T_c$ and $T_h$ under different relative humidity level (black squares for 30–35% RH and red circles for 45–50% RH) at a constant applied voltage of 0.6V.

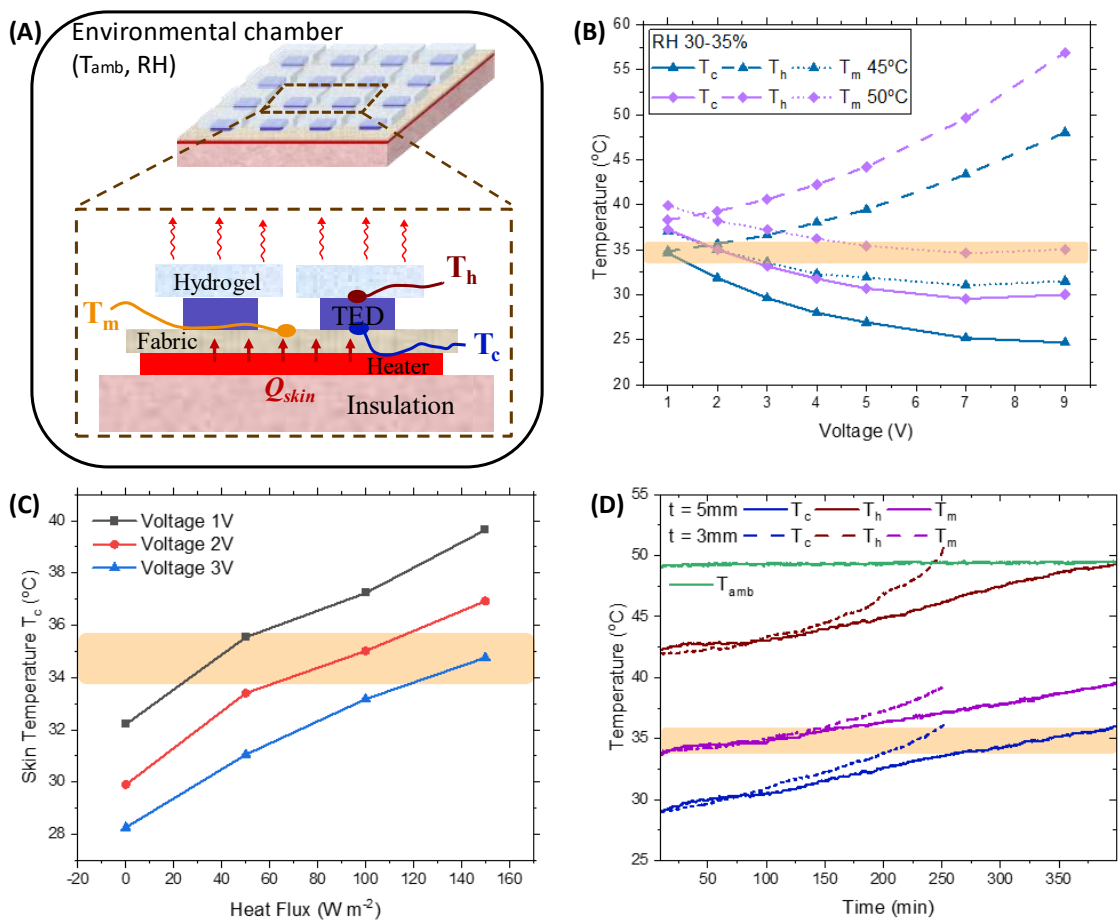

**Fig. 5. Performance of the TED-hydrogel personal cooling garment.** (**A**), Schematic diagram of the 4 × 4 TED-hydrogel personal cooling garment test setup. (**B**), Measured values of $T_c$, $T_h$, and $T_m$ (the temperature at the middle of the fabric) under different TED voltage levels and ambient temperatures in 30–35% RH and environments. The heater heat flux (*q*) is set to 100 W m$^{-2}$. (**C**), Skin temperature $T_c$ at 50°C ambient temperature and 30–35% RH, under varying human body heat flux levels. The voltage applied to the entire TED array in the garment varied from 1 to 3V. (**D**), Long-term test results for hydrogels with different thicknesses (dashed line for 3 mm and solid line for 5 mm) at 50°C ambient temperature, 30–

35% RH, and $q = 100$ W m$^{-2}$ with a fixed applied voltage to the TEDs of 5V. The shaded area represents the comfortable temperature range for the human back (33.8–35.8°C) (*17*).

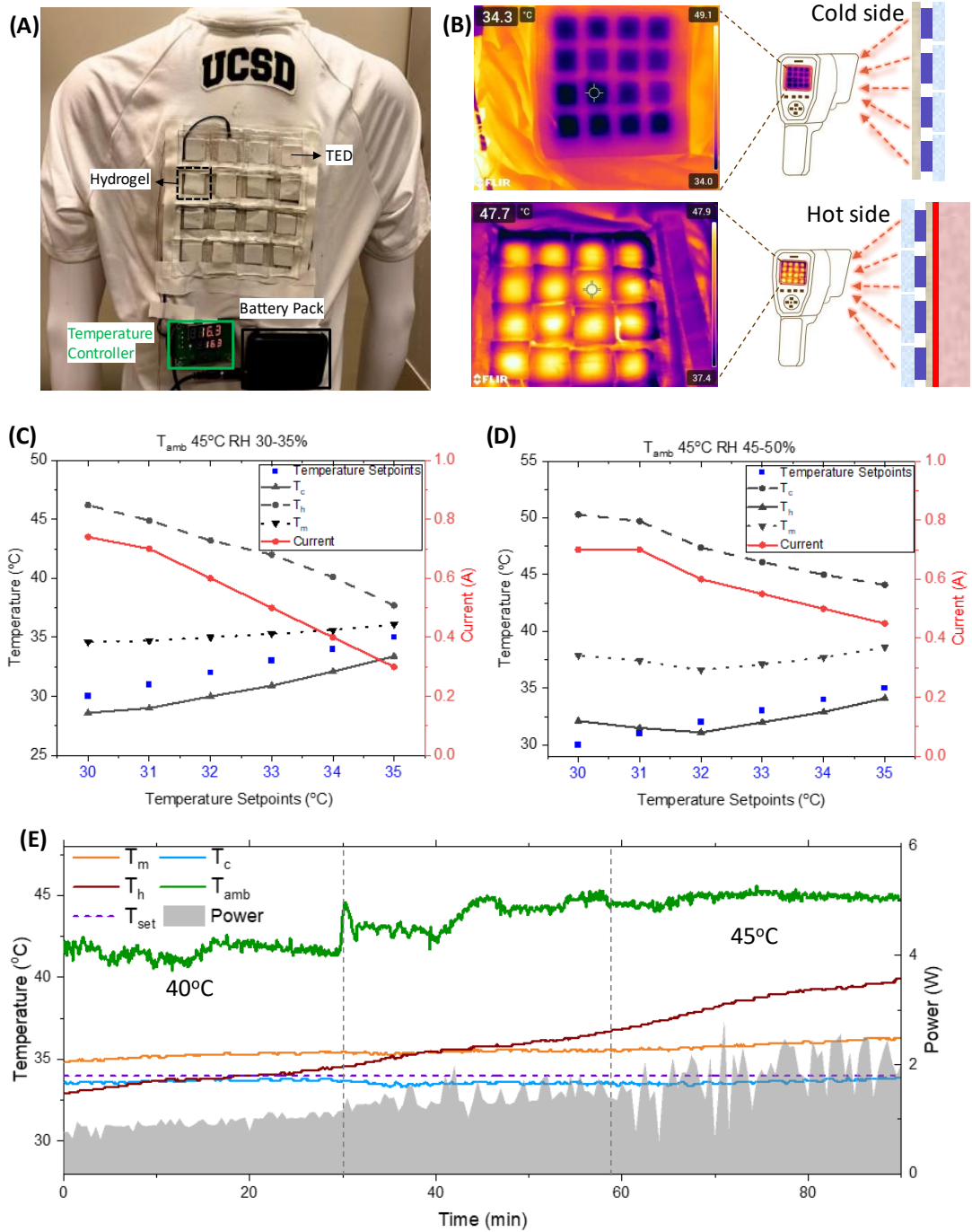

**Fig. 6. Temperature-controlled wearable TED-hydrogel device.** (**A**), Photograph of the TED-hydrogel personal cooling garment equipped with a temperature controller and a battery pack. (**B**), IR images of the cold side (taken from the fabric attached to the cold side of the TED) and the hydrogel side (taken from the top surface of the hydrogel) of the TED-hydrogel device, with ambient temperature ($T_{amb}$) at 45°C and RH at 30–35%. TED-hydrogel device response under different set temperatures at 45°C in (**C**), 30–35% RH and (**D**), 45–50% RH. The blue dots

represent the set temperature by the controller. Solid, dashed, and dotted lines represent $T_c$, $T_h$, and $T_m$, respectively. The red dots correspond to the right y-axis, representing the current input to the TEDs from the controller. (**E**), Test results for the TED-hydrogel device when the ambient temperature changed from 40°C to 45°C at 45–50% RH. The gray shaded area corresponds to the right y-axis, representing the power input to the TED. All tests in this figure used a 100 W m$^{-2}$ heat flux for the heater.

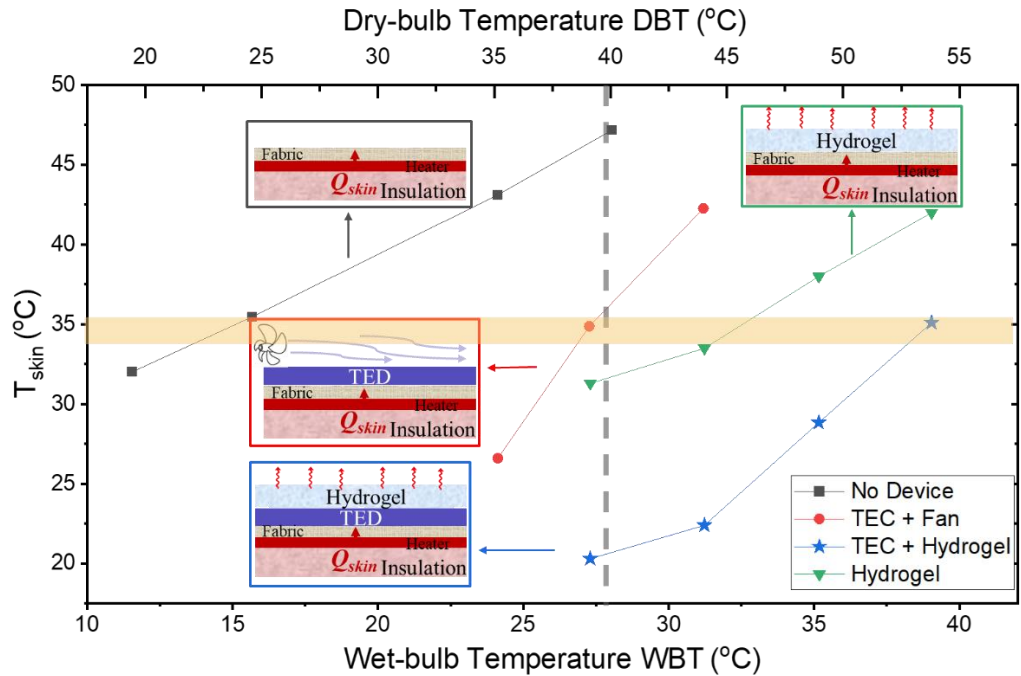

**Fig. 7. Cooling performance of the TED with forced convection of air (red circle), hydrogel alone (green triangle), and the TED-hydrogel combination (blue star).** The black line and curve square dots represent the fabric temperature variation with ambient temperature without any cooling device (No Device). The data for TED with forced convection of air and without any cooling device is from our previous work(*46*). The shaded area represents the comfortable temperature range for the human back (33.8–35.8°C)(*17*). The top and bottom x-axis represents the DBT and WBT respectively. The RH for all tests was maintained at approximately 35%.

# Supplementary Materials for

## Thermoelectrically Elevated Hydrogel Evaporation for Personal Cooling under Extreme Heat

Yu Pei *et al.*

*Corresponding author. Email: Shengqiang Cai, s3cai@ucsd.edu; Renkun Chen, rkchen@ucsd.edu

**This PDF file includes:**

Supplementary Text
Figs. S1 to S9

**Supplementary Text**

Evaporation rate calculation

The amount of evaporated water $g_s$ [kg s$^{-1}$ m$^{-2}$] can be described by the equation:

$$g_s = (25 + 19v) * (X_s - X_{sair}) * 3600 \tag{1}$$

where $v$ is the air velocity above the water, which is around 0.1 m s$^{-1}$ in our test environment, $X_s$ is the maximum humidity ratio of saturated air, $X_{sair}$ is the humidity ratio for air.

Based on the ideal gas law the humidity ratio can be expressed as:

$$X = 0.62198 \frac{P_w}{P_a - P_w} \tag{2}$$

where $P_w$ is partial pressure of water vapor in moist air, $P_a$ is the atmospheric pressure of the moist air.

The maximum amount of water vapor in the air is achieved when $P_w$ reaches the saturation pressure $P_{w,s}$ of water vapor at water surface temperature. The maximum humidity ratio of saturated air $X_s$ can be calculated by:

$$X_s = 0.62198 \frac{P_{w,s}}{P_a - P_{w,s}} \tag{3}$$

Saturated water maximum vapor pressure $P_{w,s}$ is a function of temperature only, which can be represented by an empirical formula:

$$P_{w,s} = e^{A + \frac{B}{T} + C \ln T + DT} \tag{4}$$

where T is the temperature in Kelvin,
A = 77.34,
B = -7235,
C = -8.2,
D = 0.005711.

Similarly, the humidity ratio for air $X_{sair}$ is:

$$X_{sair} = 0.62198 * \frac{RH * P_{w,s,air}}{P_a - RH * P_{w,s,air}} \tag{5}$$

where RH is the relative humidity, $P_{w,s,air}$ is the maximum saturation pressure of the water vapor at ambient air temperature.

The mass loss ratio $\frac{m_0-m}{m_0}$ can be calculated by

$$\frac{m_0-m}{m_0} = \frac{g_s \cdot t \cdot 60 \cdot A}{m_0} \tag{6}$$

where $t$ [min] is the time, and A [m²] is the surface area of hydrogel.

The heat flux $q$ [W m⁻²] corresponding to each evaporation rate can be expressed as:

$$q = g_s \cdot L_v \tag{7}$$

where $L_v$ is the latent heat of water, $L_v = 2.42 \cdot 10^6$ [J kg⁻¹] at 40°C.

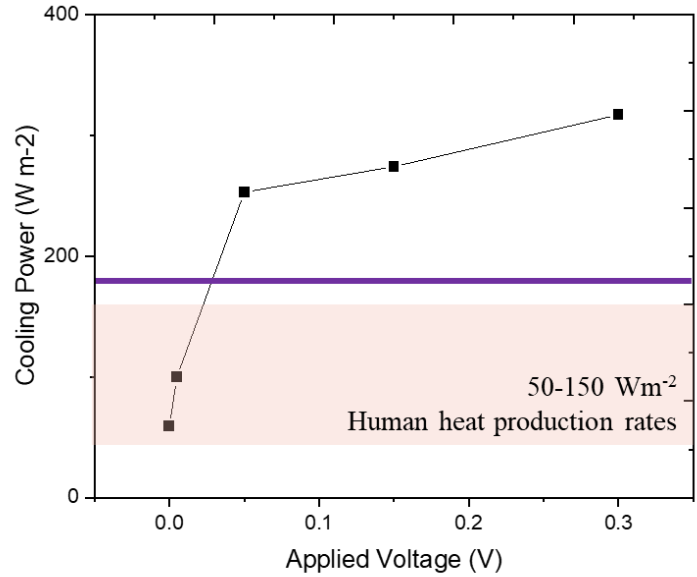

**Fig. S1.** COMSOL simulation results showing net cooling power as a function of applied voltages for hydrogel + TED device at DWT = 40°C & RH = 40% (WBT = 28.5°C). The shaded band represents the metabolic rate. Net cooling includes evaporative cooling minus the heat transfer rate from ambient to skin, with skin temperature fixed at 34°C.

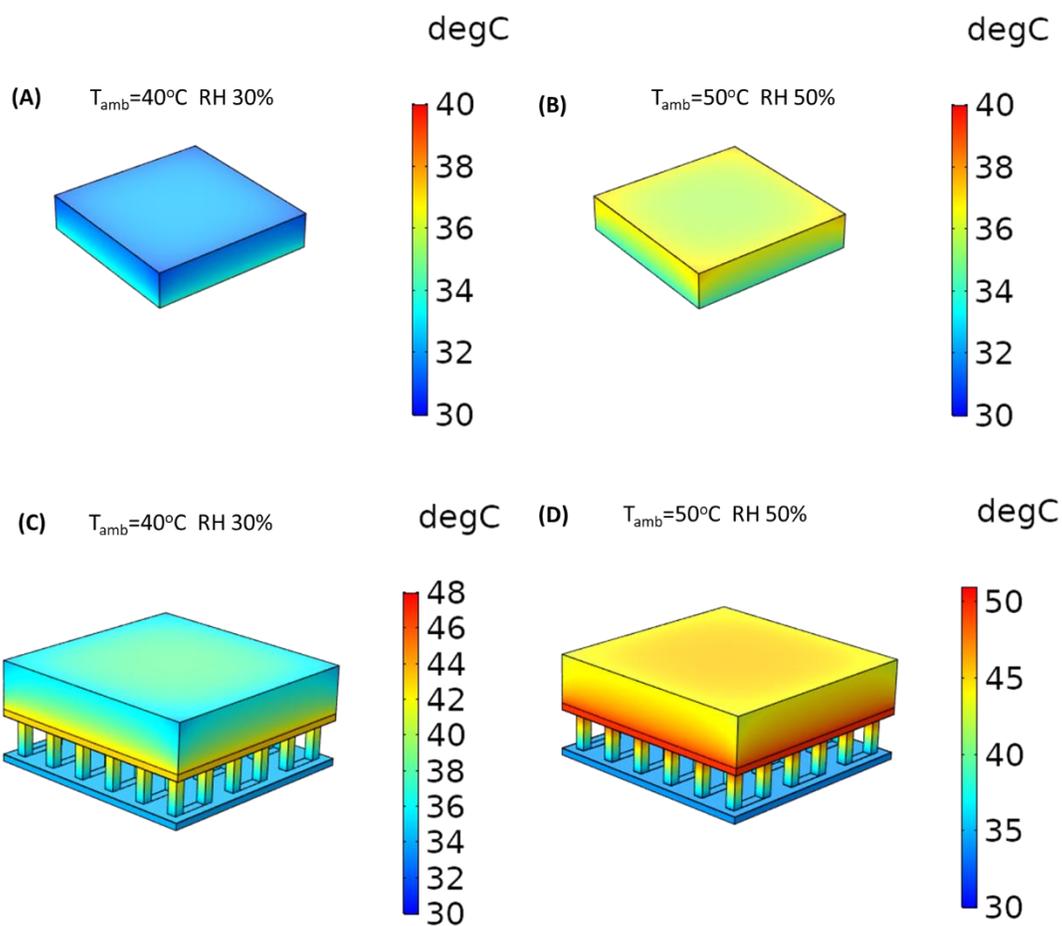

**Fig. S2.** COMSOL simulation result of temperature distribution in hydrogel. Temperature distribution for hydrogel at (A) 40 °C ambient temperature and 30% RH; (B) 50 °C ambient temperature and 50% RH. Temperature distribution for TED-hydrogel device at (C) 40 °C ambient temperature and 30% RH; (D) 50 °C ambient temperature and 50% RH.

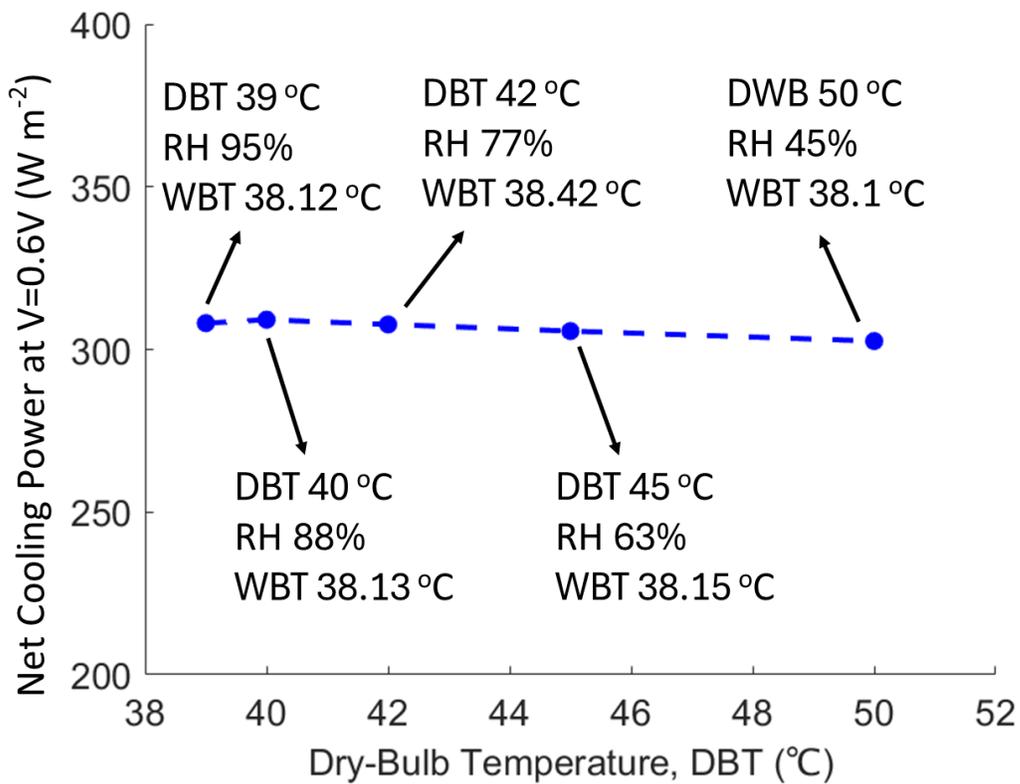

**Fig. S3.** Simulation results for the TED-hydrogel device under the same wet-bulb temperature. Modeled net cooling power of TED+Hydrogel systems as a function of dry-bulb temperature (DBT) under various relative humidity (RH) values, all corresponding to the same wet-bulb temperature (WBT ~38°C). The figure demonstrates that the TED+Hydrogel system provides the same cooling performance for a given WBT and can deliver adequate cooling even at an extreme WBT of ~38°C, regardless of DBT and RH combinations. The voltage applied to TED is 0.6V in the model. The skin temperature is set as 34 °C and heat dissipation rate from skin is 100 W m$^{-2}$.

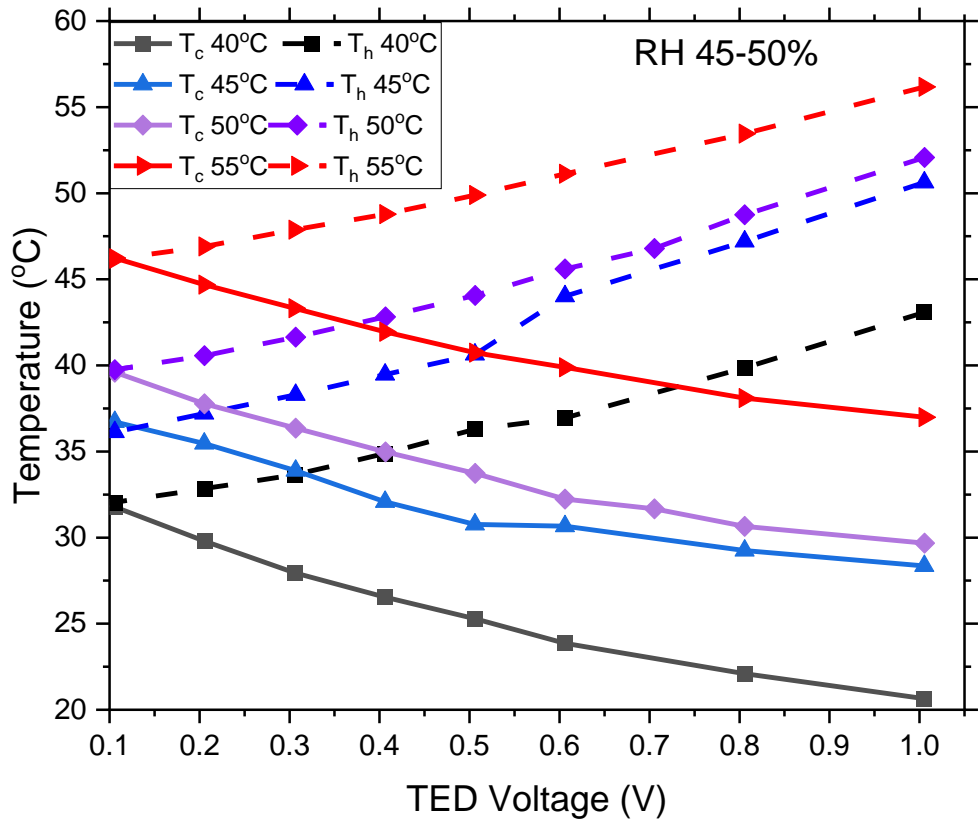

**Fig. S4.** Experiment results for the cold and hot sides of the TED-hydrogel device under different ambient temperatures and 30–35% RH. Solid lines represent the cold side ($T_c$) and dashed lines represent and hot side ($T_h$) of the TED at various TED voltages. The power supply delivers a heat flux of 100 W m$^{-2}$ to the heater to simulate human metabolic heat flux.

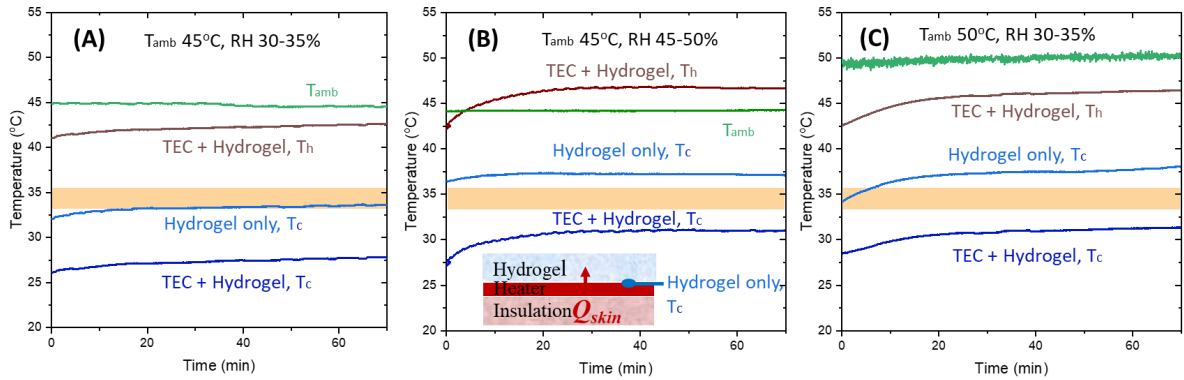

**Fig. S5**. Long-term tests for the hydrogel alone and TED-hydrogel device at a constant applied voltage of 0.6V under varying temperature and humidity conditions: (A), $T_{amb}$ = 45°C, RH 30-35%. (B), $T_{amb}$ = 45°C, RH 45-50%. (C), $T_{amb}$ = 50°C, RH 30-35%. The inset in (B) shows the schematic setup for testing hydrogel alone, with $T_c$ (of hydrogel only) indicating the temperature at the hydrogel bottom surface in contact with the heater. The shaded area represents the comfortable temperature range for the human back (33.8–35.8°C) (*17*).

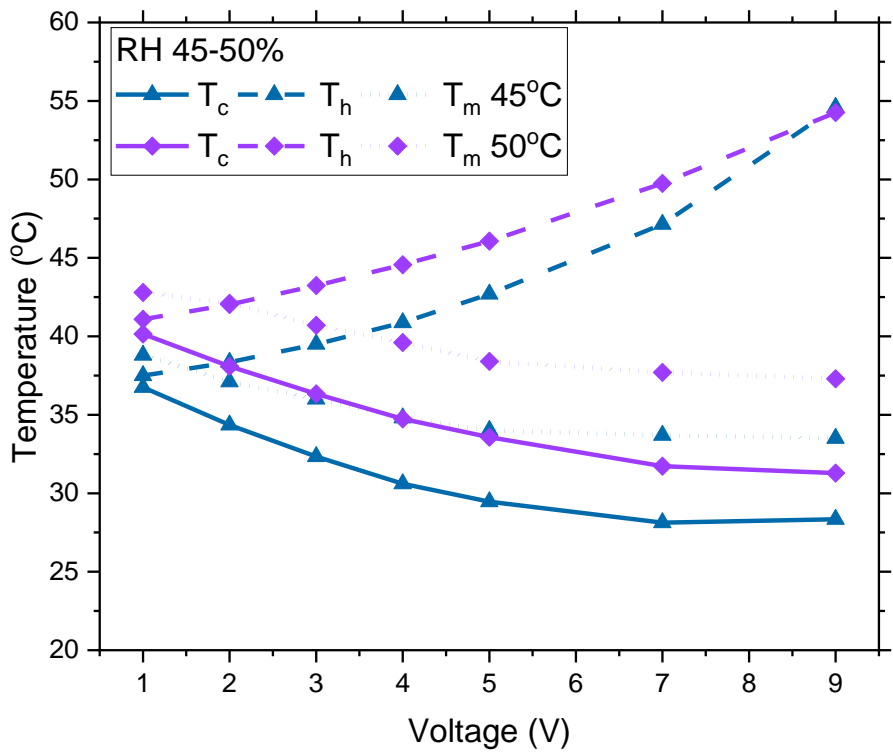

**Fig. S6**. Measured values of $T_c$, $T_h$, and $T_m$ (the temperature at the center of the fabric) under different TED voltage levels and ambient temperatures in 45–50% RH environments. The heater heat flux (*q*) is set to 100 W m$^{-2}$.

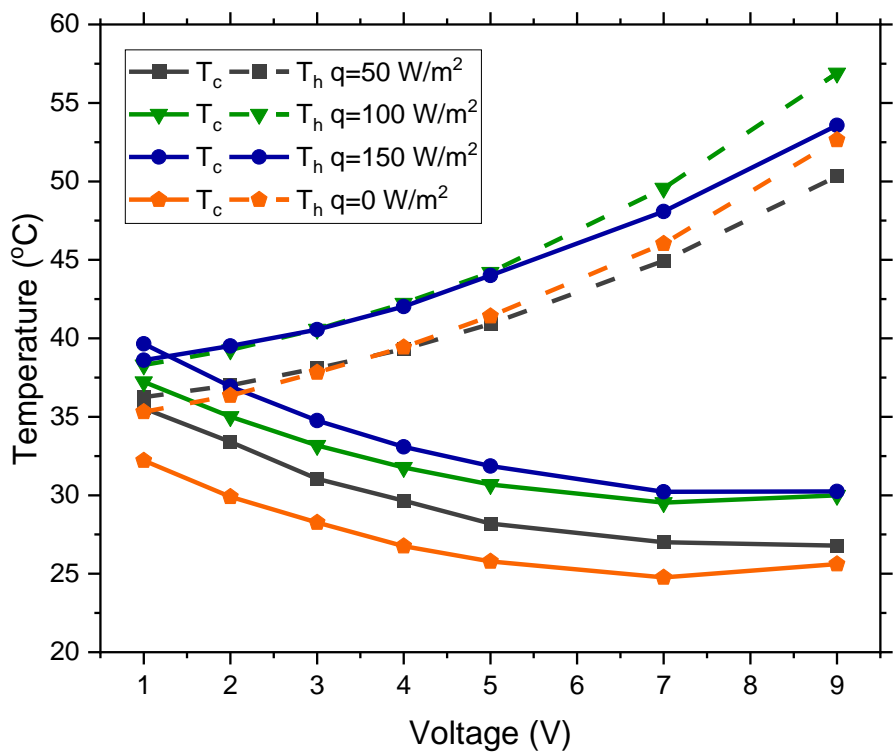

**Fig. S7**. Skin temperature $T_c$ and hot side $T_h$ of the TED-hydrogel device at 50°C ambient temperature and 30–35% RH, under varying human body heat flux levels. The voltage applied to the entire TED array in the garment varied from 1 to 9V.

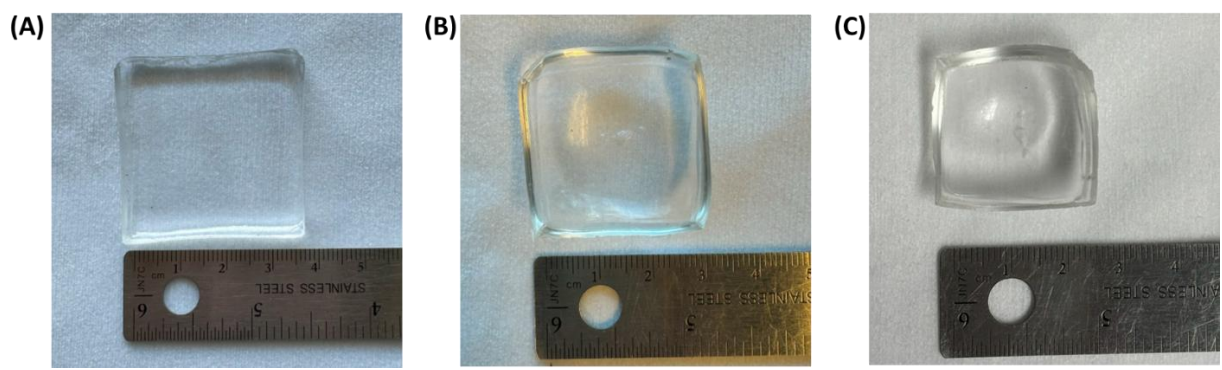

**Fig. S8.** Changes in 5mm thick hydrogel size over test time. (A) Before test (side length ~3.7cm). (B) After 2 hr test (side length ~3cm). (C) After 4 hr test (side length ~2.5cm). Test environment: ambient temperature 50 °C, humidity 30%-35%.

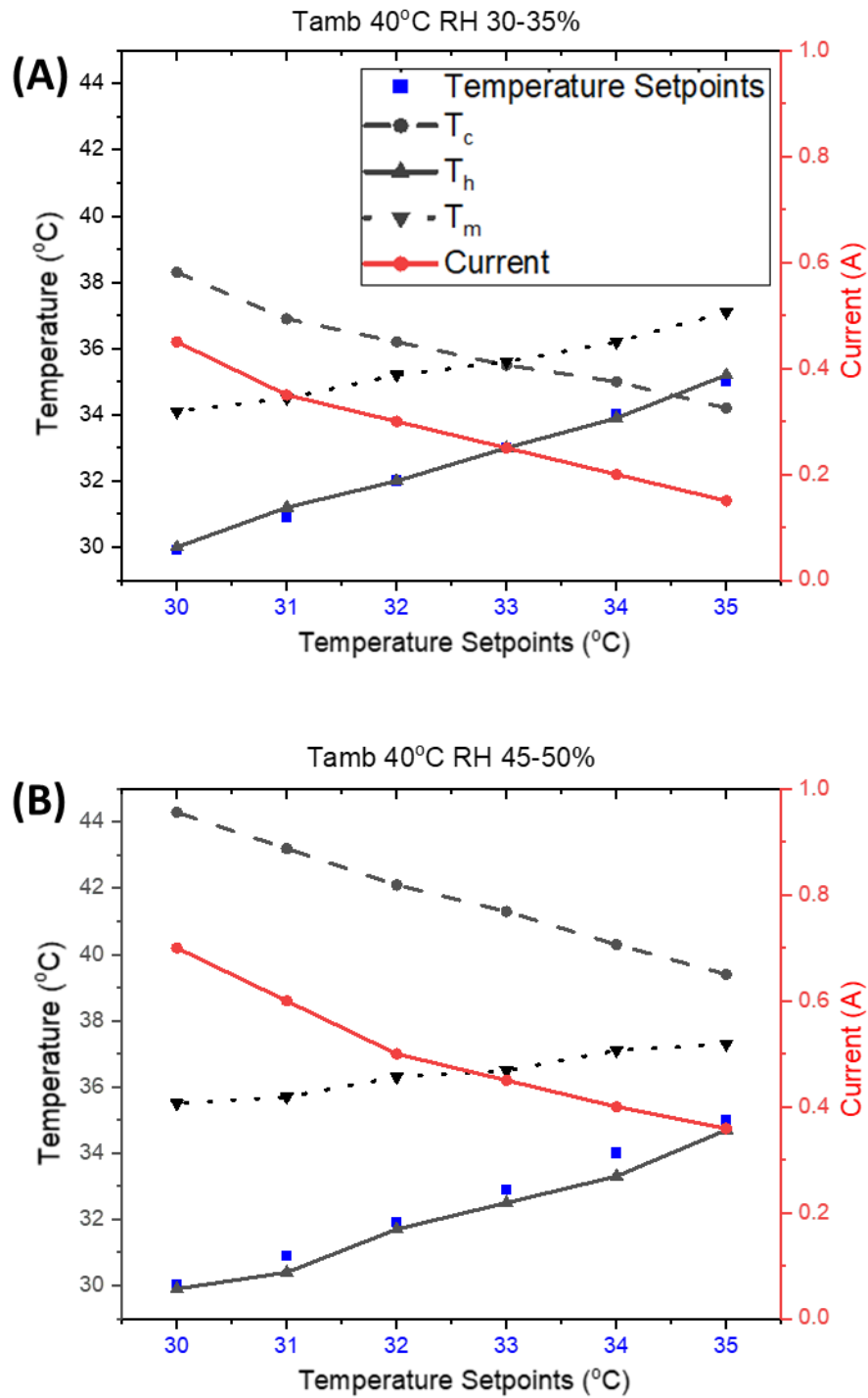

**Fig. S9.** TED-hydrogel garment temperature response under different setting temperature. Ambient temperature at 40°C and relative humidity is (A) 30–35% RH and (B) 45–50% RH. Hydrogel thickness is 5mm. Heat flux is 100 Wm$^{-2}$.

12